\documentclass[12pt]{article}
\usepackage{amsmath,amsfonts}
\usepackage{amscd,amsthm}
\usepackage{geometry}
\usepackage{graphicx}
\usepackage{color}
\usepackage[applemac]{inputenc}
\usepackage{tikz}
\usepackage{graphicx}
\usepackage{color}
\usepackage[applemac]{inputenc}
\usepackage{natbib}
\usepackage{cases}
\usepackage{tikz}
\usepackage{algorithm2e}
\usepackage{setspace} 

\begin{document}

\title{Parallel versus Sequential Update and the Evolution of Cooperation with the Assistance of Emotional Strategies}
\author{Simone Righi\thanks{MTA TK "Lend\"{u}let" Research Center for Educational and Network Studies (RECENS),
Hungarian Academy of Sciences. Mailing address:
Orsz\'{a}gh\'{a}z utca 30, 
1014 Budapest, Hungary.
Email: simone.righi@tk.mta.hu.} \and K\'{a}roly Tak\'{a}cs \thanks{MTA TK "Lend\"{u}let" Research Center for Educational and Network Studies (RECENS),
Hungarian Academy of Sciences. Mailing address:
Orsz\'{a}gh\'{a}z utca 30, 
1014 Budapest, Hungary.
Email: takacs.karoly@tk.mta.hu}}
\date{January 19, 2013}
\maketitle
\begin{abstract}
Our study contributes to the debate on the evolution of cooperation in the single-shot Prisoner's Dilemma (PD) played on networks. We construct a model in which individuals are connected with positive and negative ties. Some agents play sign-dependent strategies that use the sign of the relation as a shorthand for determining appropriate action toward the opponent. In the context of our model in which network topology, agent strategic types and relational signs coevolve, the presence of sign-dependent strategies catalyzes the evolution of cooperation. We highlight how the success of cooperation depends on a crucial aspect of implementation: whether we apply parallel or sequential strategy update. Parallel updating, with averaging of payoffs across interactions in the social neighborhood, supports cooperation in a much wider set of parameter values than sequential updating. Our results cast doubts about the realism and generalizability of models that claim to explain the evolution of cooperation but implicitly assume parallel updating. 
\vspace{1cm}

\end{abstract}

\newpage

\setlength{\baselineskip}{20pt}
\singlespacing

\section{Introduction}\label{Introduction}
The defining characteristics of Complex Adaptive Systems (CAS) is that they are made of a large number of interacting components that - together - generate results which are not observable at the level of each single element (\citealt{anderson1972more}). In such systems, even relatively simple local interaction rules can result in very complex behaviors of the aggregate system (\citealt{helbing2012social}). Moreover, small variations in the local rules of interaction can result in large (non-linear) changes at the system scale. Thus, when studying complex systems, it is important to understand the impact of changes in the rules of interaction on the aggregate system properties. 

One of the most elaborate complex system known is the human society. The human society is composed of individuals who are already complex in themselves and who interact with each other in highly complex ways and patterns. The results of these interactions are emerging structures, whose behavior can hardly fit simple or linear models (consider, for instance, financial markets and traffic in cities that are well studied examples). 

What makes social systems uniquely complex is that their components are self-aware and, as such, act with a certain degree of intentionality. Game theory, that first attempted to model formally the complexity of interdependent intentional decisions (\citealt{von1944game}), made drastic assumptions initially that (1) humans act rationally and (2) small scale interactions can be aggregated to the system level through simple extrapolation. These assumptions, however, have been relaxed progressively and now social systems are studied considering individuals that lack perfect foresight about the future consequences of their actions (\citealt{march1978bounded,simon1982models}) and are affected by emotions and feelings (\citealt{camerer2003behavioral,gigerenzer2008gut}). In this context, an interesting approach to the problem of studying social interactions  is provided by evolutionary game theory that tries to explain which strategies disappear, survive or thrive in the long run where strategies with higher payoffs tend to diffuse. This strain of literature allows to identify reasons and situations in which strategies that are rational in static games, are not the most successful ones in an evolutionary context. One of the problems to which this literature has been applied is the theoretical justification of the continuing  existence of selfless cooperative behavior in both nature and society. 

The survival and extent of cooperative behavior in human society has for a long time been considered as one of the main and most difficult questions in social science (\citealt{axelrod1981evolution,axelrod1997complexity,axelrod2006evolution}). Social dilemma games describe situations in which the self-interest of agents is in contrast with the one of their interaction partners. The most studied and puzzling among them is the Prisoner's Dilemma (PD). Individuals have two options in the PD: the dominant strategy - defection - guarantees a higher payoff regardless of what the partner does, but the alternative strategy - cooperation -  if played mutually, offers a payoff that is higher than the payoff from mutually playing the dominant strategy. It has been shown that unstructured populations with individuals interacting with randomly selected partners are unable to solve the puzzle of cooperation as natural selection generates uniform populations of defectors (\citealt{taylor1978evolutionary,hofbauer1998evolutionary}). Recently, particular interest was devoted to the issue of evolutionary games in structured (networked) populations. On the one hand, studying interactions on networks increases significantly the realism of models, as this formalism allows to explicitly consider their inherent locality. On the other hand, limiting the possible interactions of agents (given the sparseness of interactions), proved to be able to increase cooperation in the population (\citealt{nakamaru1997evolution,nowak1992evolutionary}).  

The structure of interactions is important because many relevant mechanisms are channeled through network ties and because behavioral influence spread differently in different structures. Similarly, reputational mechanisms such as image scoring (\citealt{wedekind2000cooperation}) also flow via network ties. It has been studied which network topologies are most efficient for the emergence and diffusion of cooperation (\citealt{Virtuallabs,santos2005scale,johnson2003social}). Having formal analytical proofs, however, is difficult in this context. Most contributions, therefore, use numerical simulations and agent-based models. This is especially true for the case in which  the co-evolution of network topology and agent types (\citealt{santos2006cooperation,yamagishi1996selective,yamagishi1994prisoner}) is studied. Among the mechanisms that improve the conditions of cooperation in dynamic networks are the possibility of parter selection, exclusion of defecting agents, and exit from relationships (\citealt{schuessler1989exit,vanberg1992rationality,yamagishi1996selective}).  

In \cite{righi2014emotional} we study the conditions for the emergence of cooperation on dynamic signed networks. Virtually all co-evolutionary models of networks and cooperation before assumed the presence of positive relations only. We have relaxed this assumption and interpret signed ties as expressing the (positive or negative) emotional content of the social relationship between two individuals. This interpretation is consistent with evidence that emotions evolved in humans due to their their function in social interactions (\citealt{darwin1965expression,frank1988passions,keltner2006social,trivers1971evolution}). Signed relations help guaranteeing the diffusion of reputational information about agent's past conduct, thus providing a guiding light for partners in choosing the correct behavioral response. While relational signs could be interpreted as a form of memory (\citealt{szolnoki2013evolution}), they constitute a cognitively much less costly mechanism, which can be used as a shorthanded tool that condenses the past history of a relationship.  
The sociological intuition behind why negative ties should also be considered for the evolution of cooperation is the relevance of altruistic punishment of defectors (\citealt{bowles2004evolution,dreber2008winners,ernst2005human,fowler2005altruistic,fowler2005egalitarian}) and the process of stigmatization and social exclusions of these individuals (\citealt{kerr2008detection,kurzban2001evolutionary}). These mechanisms could result in negative interpersonal ties that in turn help the spread of cooperative behavior.

In \cite{righi2014emotional} we thus construct an evolutionary model where agents play the Prisoner's Dilemma on signed networks (where links can be either positive or negative). We assume that tense relationships can be resolved either by changing the sign or being erased and rewired. We analyze a setup in which network topology co-evolves with relational signs and agent strategies. Our major conclusion is that the introduction of conditional strategies, that utilize the emotional content embedded in network signs, can act as catalysts and in general create favorable conditions for the spread of unconditional cooperation. We notice, however, that the introduction of conditional strategies is successful in eliciting increased cooperation only if the network is dynamic (i.e. only if there is some positive probability of updating the network topology). Our results are summarized in Table \ref{SummaryResults}.\footnote{This table and the results proposed for the parallel updating case are taken from \cite{righi2014emotional}.} In line with the literature, we find that the evolution of unconditional cooperation occurs most likely in networks with relatively high chances of rewiring and low likelihood of strategy adoption (or strategy evolution). While some rewiring enhances cooperation, too much rewiring limits its diffusion. Finally, we provide evidence that, unlike in  networks with positive ties only, cooperation becomes more prevalent in denser networks. 

\begin{table}[!ht]
\centering
\begin{tabular}{| p{4cm} | p{4cm} | p{4cm} |}
\hline 
 & Without rewiring & With rewiring \\
\hline 
Without emotional strategies & No cooperation & Cooperation through clustering of strategies \\
\hline 
With emotional strategies & Some cooperation, only if most agents are emotional  & {\bf The emergence and diffusion of cooperation} \\  
\hline 
\end{tabular}
\label{SummaryResults}
\caption{Summary of main findings in \cite{righi2014emotional}. A positive rewiring probability and the initial presence of conditional strategies are both required for the evolution of cooperation.}
\end{table}

In the present study we use this general setup to analyze the influence of the update rule, which essentially defines how and when agents interact, on the chances of cooperative behavior to become widespread in the population. In this way, our research follows the pathway of earlier studies that examined synchronous vs real-time interactions in social dilemmas (\citealt{huberman1993evolutionary}). Two different types of updating are proposed. The first is {\it sequential}, where single couples of agents are selected for interaction and as a consequence, the payoffs obtained in their interaction drive evolutionary and network update. The second is {\it parallel}, in which all agents play at the same time and average payoffs from the interactions with neighbors are calculated and used to drive the evolutionary process. 

We show how the survival and diffusion chances of cooperation depend strictly on the type of updating rule used. We provide evidence that under a rather general set of parameters combinations, the parallel updating rule provides better conditions for the diffusion of cooperation as it allows conditional strategies that makes use of emotions to act as catalyst of virtuous behavior.  Where the sequential update is applied instead, unconditional defection progressively diffuses and comes to dominate the population.

The remaining of the paper is divided as follows. In the next section, we describe our model and its characteristics as they were proposed in  \cite{righi2014emotional}. In addition, we describe the details of the two updating rules that we study. The following Section \ref{Results} presents our new results, while a discussion concludes (Section \ref{concl}).

\section{The Model}\label{TheModel}

We consider the model first introduced in \cite{righi2014emotional}. We study a population of $N$ agents. Agents are connected initially in a random network (\citealt{erdHos1959random})  where each possible edge exists with probability $\rho\in[0,1]$.  The cardinality $k_i$ of $\mathcal{F}_i$ is the degree (or number of network contacts) of agent $i$. The network is signed and each network tie is labelled either \textit{negative} or \textit{positive}. In this paper, we report results from setups where each link is initialized with the same probability ($1/2$) as either negative or positive.

Each agent in the population can interact and play the single-shot Prisoner's Dilemma (with binary options of cooperation or defection) with partners selected from its first order social neighborhood. Among the social dilemmas, the Prisoner's Dilemma is the one that sets the stakes the most against the emergence of cooperation since it is characterized by the classical payoff structure Temptation(T) $>$ Reward(R) $> $ Punishment(P) $> $ Sucker(S) (Table \ref{PDPayoffs}). 

\begin{table}[!ht]
\centering
\begin{tabular}{|r|c|c|}
\hline
& \textrm{C} & \textrm{D} \\
\hline 
\textrm{C} &  $(R=3,R=3)$& $(T=5,S=0)$ \\
\hline 
\textrm{D} & $(S=0,T=5)$ & $(P=1,P=1)$ \\
\hline
\end{tabular}
\label{PDPayoffs}
\caption{The Prisoner's Dilemma payoff matrix. The numerical payoffs used here are the same of \cite{axelrod2006evolution}}
\end{table} 

As discussed, we assume that network signs, embedding an emotional content that follows from previous interaction, can affect behavior. From this point of view, we can characterize three types of strategies:
\begin{itemize}
\item Unconditional Defection (UD);
\item Unconditional Cooperation (UC); 
\item Conditional Strategy (COND): cooperate if the tie with the interaction partner is positive; and defect otherwise.
\end{itemize}
While UC always cooperates and UD always defects, the strategy COND is conditional on the sign of the link between the interaction partners. The COND strategy, therefore, can be interpreted as a differentiated emotional reaction or affectional response towards others with harmonic or disharmonic interaction record as it prescribes cooperation with agents connected with a positive tie, and defection with partners connected with a negative relation. Below, we report results for initialization in which agents are assigned with one of the three strategies randomly in equal proportions: $(1/3=\mu_{UC}=\mu_{UD}=\mu_{COND}=1/3)$. 

As discussed, our model allows for the co-evolution of network signs, agent strategies and network topology. Network signs and agent behavior influence each other and the latter also affects the evolution of network topology. Each of these modules requires some clarification. 

\paragraph{Sign update: agent behavior influences relational signs.} Relational sign update simulates the consequences of behavior on the emotional relationship with peers. This type of update happens automatically, i.e. it is not part of individual strategies. It is relatively straightforward to assume that a relationship in which both agents defect turns negative and one in which both agents cooperate turns positive. When actions differ, we have a more complex case. In this situation an asymmetric tension arises since the cooperator could be frustrated of having a positive tie with a defector and the defector could be shamed of its action towards someone who shares a positive sign with him. In this case we assume that the link {\it could} change its sign. Specifically, we assume that a frustrated positive link can turn negative with probability $P_{neg}$ and a frustrated negative link can turn positive with probability $P_{pos}$. Given the payoff structure of the PD (where a cooperator always obtains a very low payoff when its partner defects obtaining a high payoff), it is logical to assume that the frustration from disappointment is larger than the frustration from shame, that is $P_{neg}>>P_{pos}$. In particular we fixed $P_{neg}=0.2$ and $P_{pos}=0.1$.

\paragraph{Network topology update: rewiring.} In addition, we allow for an endogenous update of network topology (as suggested, for example by \citealt{santos2006cooperation}). It means that behavior can influence the network structure directly. With a probability $P_{rew}$ (named {\it rewiring probability}) the frustration, emerged as a consequence of  different strategies played in the PD game, can lead the frustrated agent to sewer its relationship and to search for a new partner. From the technical point of view, the rewiring assumes a certain degree of  transitive closure (\citealt{granovetter1973strength}), meaning that we allow new connections to be created only between friends of friends. This modeling choice naturally follows from sociological observations.  Still, with a small but positive probability (fixed in the following to $P_{rand}=0.01$) the new link can be constructed with a randomly selected new partner. 

\paragraph{Strategy update: parallel vs dyadic update.} Individual payoffs measure the efficiency of an agent's strategy in its social neighborhood. In this paper, we focus on how the choice of the timing of this update changes the results regarding the emergence of cooperation in signed networks. In particular, two alternative types of updating rules are studied:
\begin{itemize}
\item {\bf Sequential Update}: At each time $t$, two connected agents are selected randomly for playing the PD. After playing and observing the relative payoffs, the agent with a strictly lower payoff (if any) updates its type and adopts the strategy of the more successful partner with probability $P_{adopt}$ (assumed to be equal for all agents). Therefore, only the current dyadic payoff matters in the determination of the survival chances of a strategy. Including the modules discussed above, Algorithm \ref{sequentialalgo} reports the pseudo-code of the intra-step dynamics with sequential updating. 

\begin{algorithm}[!ht]
\SetAlgoLined 
 Select randomly two connected agents ($i$ and $j$)\;
 Play the PD and compute payoffs\;
 Update relational signs between $i$ and $j$\;
 \If {link is tense}
 {
 Rewire link between $i$ and $j$ (with probability $P_{rew}$)\; 	
 }
\If {link is tense and not rewired} 
{
The agent with (strictly) lower payoff adopts the strategy of the partner (with probability $P_{adopt}$)\; 
}

\caption{Intra-step dynamics, repeated at each time step $t$, in the sequential update case.}
\label{sequentialalgo}
\end{algorithm}

\item {\bf Parallel Update}: At each time $t$, for each agent $i$ the average payoff across all its interactions is calculated.  Each agent then compares its payoff with the one of all peers in its first order social neighborhood. If a subset of these agents has a payoff higher than its own, then agent $i$ will adopt the strategy played by one of them, selected uniformly at random. Evolutionary update happens, for each agent, with probability $P_{adopt}$, which is assumed to be equal for all players. In order to avoid that the order in which we select the agents influences the outcome, each of them refers to the situation at $t-1$ when changing either its relational signs or the network topology at time $t$. Moreover, the update of strategies happens for each agent after observing payoffs, at time $t$, of every other agent. Again, including the modules discussed above,  Algorithm \ref{parallelalgo} reports the pseudo-code for our model intra-step dynamics with parallel update.
\begin{algorithm}[!ht]
 \SetAlgoLined 
 \For{each agent $i$}
 {
 	Compute its social neighborhood $\mathcal{F}_i^{t-1} \in N$\;	
	 \For{each agent $j \in \mathcal{F}_i^{t-1}$}
	 {
		Play the PD and compute payoffs\;
		Update relational signs between $i$ and $j$\;
		Rewire link between $i$ and $j$ if tense (with Probability $P_{rew}$)\; 
	}	
	Compute average payoff of agent $i$\; 
}	
\For{each agent $i$}
{
  	Observe the average payoffs of each agent $j \in {F}_i^{t}$\; 
	Adopt a random (strictly) better strategy (with probability $P_{adopt}$)\; 
}
\caption{Intra-step dynamics, repeated at each time step $t$, in the parallel update case.}
\label{parallelalgo}
\end{algorithm}
\end{itemize}

One can immediately appreciate that the sequence of events is identical in the two implementations. The number of agents that play at each time step and the rule used to determine the evolutionary update, however, are different. While these differences seem minimal, they are consequential for the chances of cooperation to evolve in dynamic signed networks.

\section{Results}\label{Results}

\subsection{Evolution with sequential and parallel updating}\label{subsecexaples}
The main objective of this paper is to figure out whether the rule of update influences the chances of emergence for cooperation in the single-shot PD played on signed networks. 
\begin{figure}[!ht]
\centering
\includegraphics[width=0.49\textwidth]{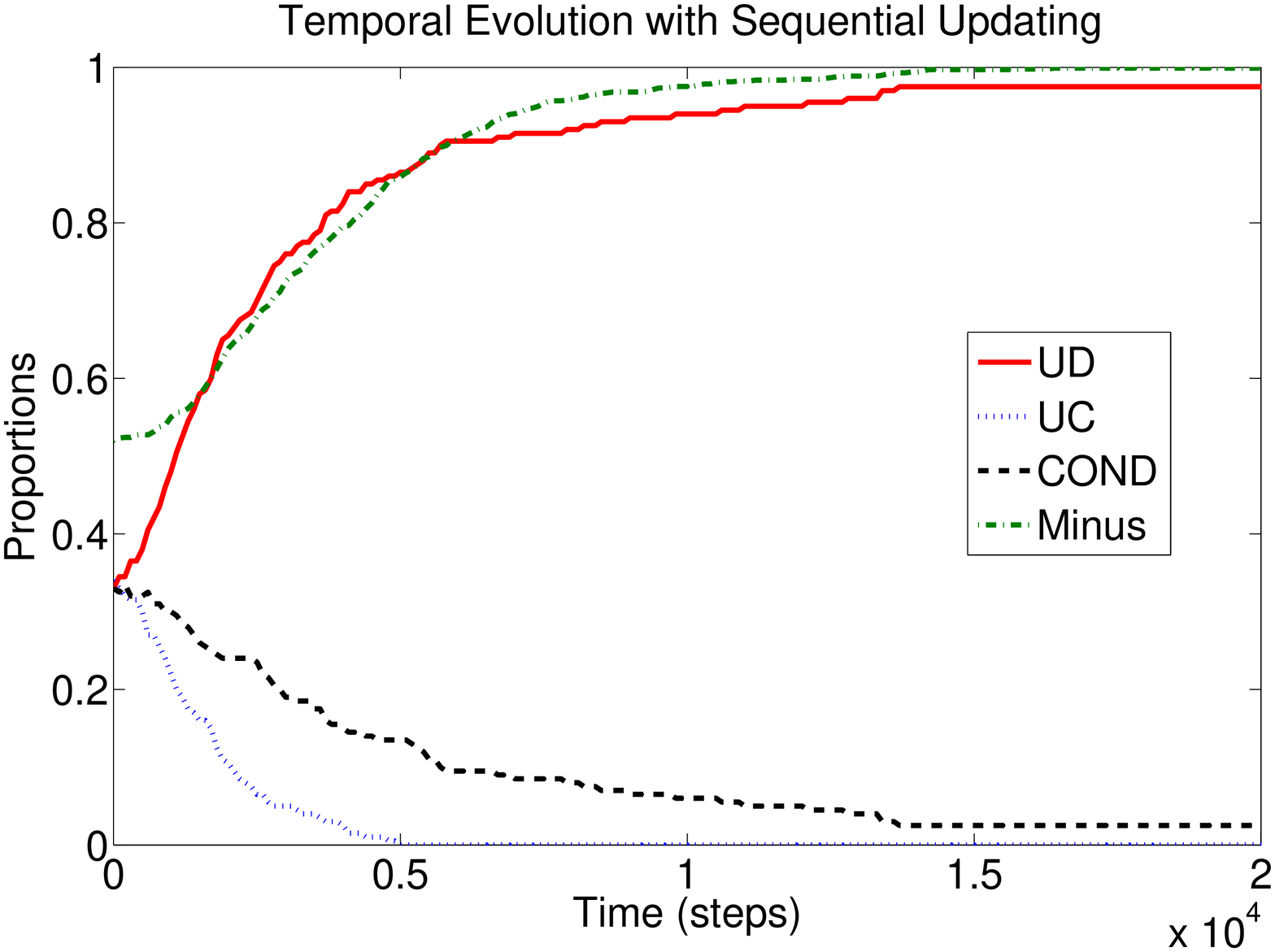}
\includegraphics[width=0.49\textwidth]{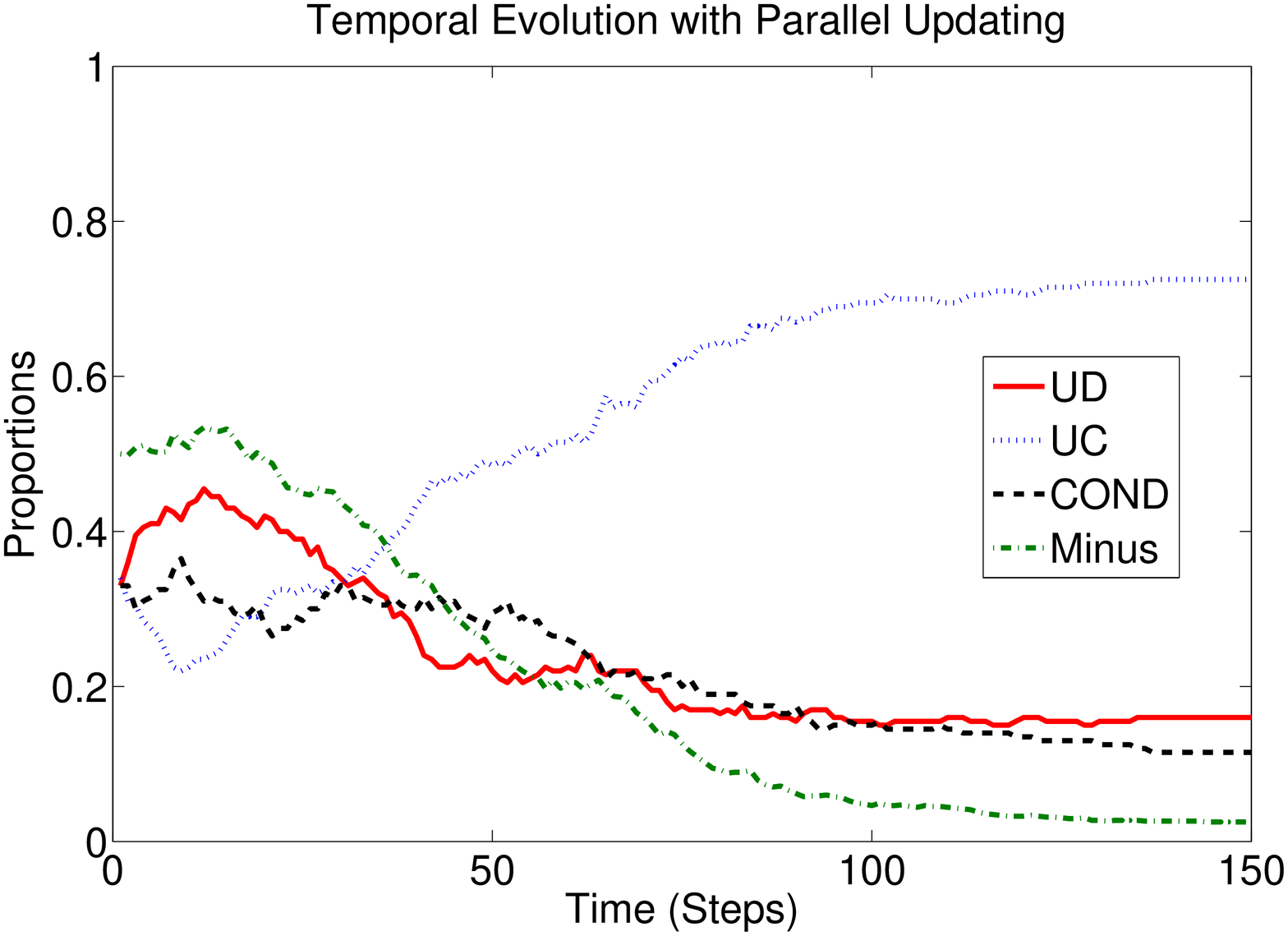}\\
\includegraphics[width=0.49\textwidth]{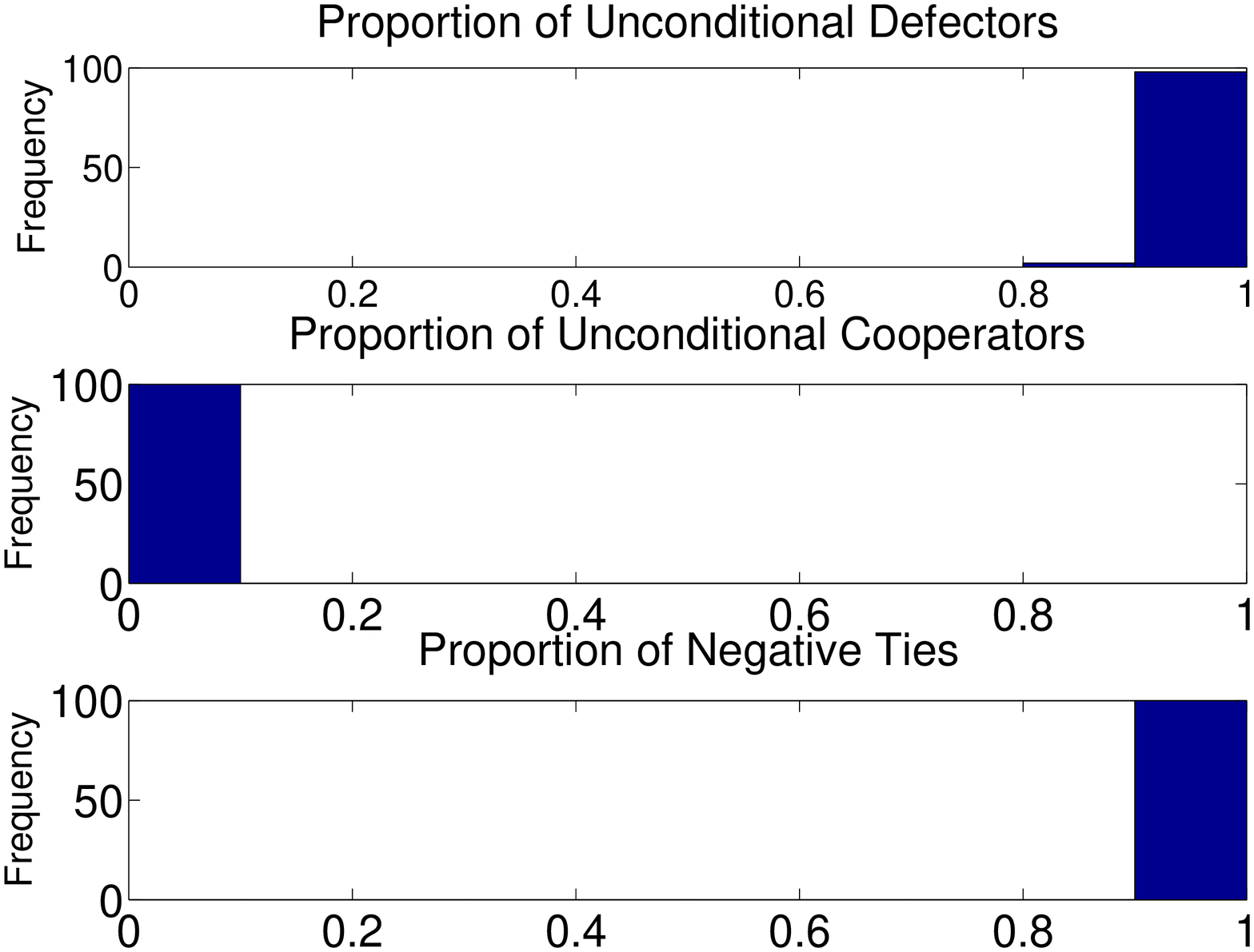}
\includegraphics[width=0.49\textwidth]{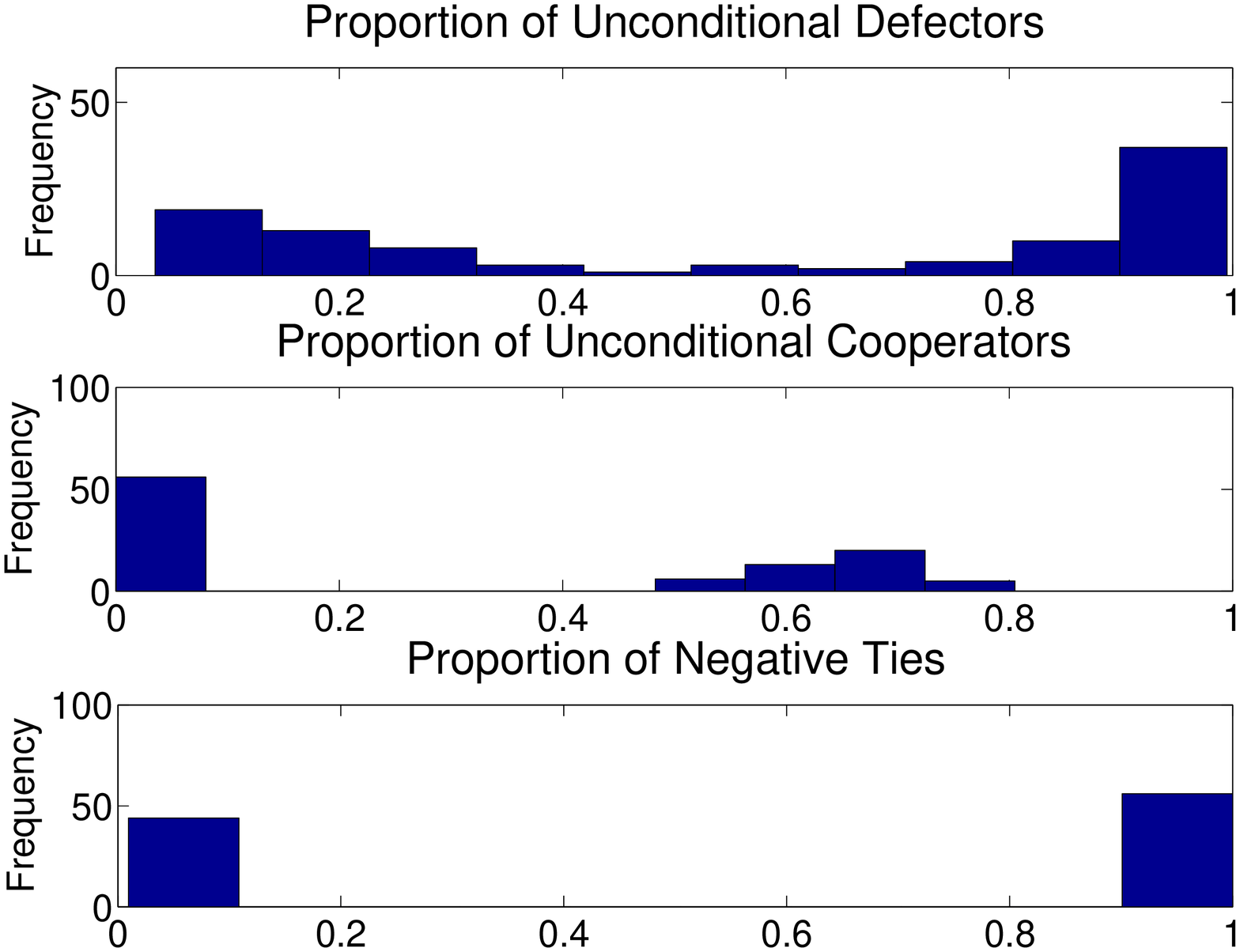}
\caption{Upper Panels: dynamic evolution of the proportions of agent types and network signs in typical simulations. The left panel shows the evolutionary process with {\it sequential updating} and the right panel with {\it parallel updating}. The lower panels depict the distribution of the final proportions of UDs, UCs, and negative ties in the sequential (left panel) and parallel (right panel) update dynamics (calculated on 100 simulations each). For all simulations: N=200 and $P_{rew}=P_{adopt}=0.1$. The initial population is divided equally among UC, UD, and COND strategies. Moreover, the network signs are randomly initialized positive or negative with equal probability and the probability of existence for each tie is $P_{link}=0.05$.}
\label{examples_and_distros}
\end{figure}

We find that the two implementations differ radically with regard to the chances of cooperation to emerge (Figure \ref{examples_and_distros}). Using the sequential update, all forms of cooperation (both in conditional and unconditional strategies) are progressively eliminated from the population and remaining strategies all defect. Indeed, while the disappearance of CONDs is slowlier than the one of UCs (due to their relatively better performance against UDs), given that all signs progressively become negative, all remaining conditional strategies act as defectors and are effectively impossible to discern them from universal defectors.
This type of evolution, whose statistical relevance for the selected parameters set is shown in the lower panel of Figure \ref{examples_and_distros}, is not limited to these conditions, and it holds in general.  This system level evolution follows from the nature of the micro-level interactions. The COND players safeguard themselves from direct exploitation from UDs by exploiting the emotional content of the relationship embedded in the link. In a dyadic comparison, however, they can never outperform the latter as they progressively diffuse in the population. As a matter of fact, while UCs are systematically exploited by UDs and thus destined to disappear rather quickly, CONDs have the "choice" of either progressively turning their links to UD players to negative, thus becoming functionally equivalent to them; or being progressively eliminated. 

The mechanism of rewiring of tense connections has been shown to help the survival of cooperation in networks (\citealt{yamagishi1996selective,yamagishi1994prisoner}). It does so by segregating agents by type and thus increasing the probability that a cooperator plays with another cooperator (\citealt{becker1976altruism,nowak2006five,nemeth2007evolution}). The level of rewiring, however, that is proposed in Figure \ref{examples_and_distros}  is not sufficient to guarantee the survival of cooperation. We will provide a more comprehensive study of the impact of this variable in the following.

When parallel update is applied, things change in favor of the emergence of cooperation. Now the UCs tend to dominate the population at the end of a significative number of simulations. COND players are able to obtain payoffs  that are higher than those obtainable by an UD in a mixed population, because averages are calculated from all interactions in the social neighborhood. While the CONDs do not gain dominance themselves,  their presence allows for the evolution of unconditional cooperation. Again, a look at the interaction level is pivotal to understand these results. The mechanism allowing this emerging behavior, described in \cite{righi2014emotional}, relies on the fact that conditional players tend to develop a collaborative relationship with UCs while not being systematically cheated by UDs. This ensures good performances of those COND players that act as interphase between the two pure strategy types. Dynamically, the UDs progressively become CONDs and these, in turns tend to become UCs (which in a connected and clustered world dominated by cooperation is the strategy with the highest average payoff). This effect is reinforced by the presence of the possibility of sewering the relationship and rewiring it with a friend of a friend as negative links can also be erased, which tends to isolate defectors from cooperators.

\subsection{The two main dynamics: adoption vs rewiring}\label{rewvsadopt}

Let's now discuss the results of the previous section in a more systematic fashion. Our model's evolution is driven by two major forces. First, agents with lower average payoff adopt strategies in their social neighborhood that perform better (adoption, or evolutionary, dynamics). Second, stressed relationships can be rewired (rewiring dynamics). In order to analyze the joint influence of these two important forces, we study their effects systematically changing their relative strength (measured as their probability to happen, respectively, at each time step and interaction). Logically, this is a similar inquiry to the analysis of network and strategy update in models with positive ties only (\citealt{santos2006cooperation}).

\begin{figure}[!ht]
\centering
\includegraphics[width=0.32\textwidth]{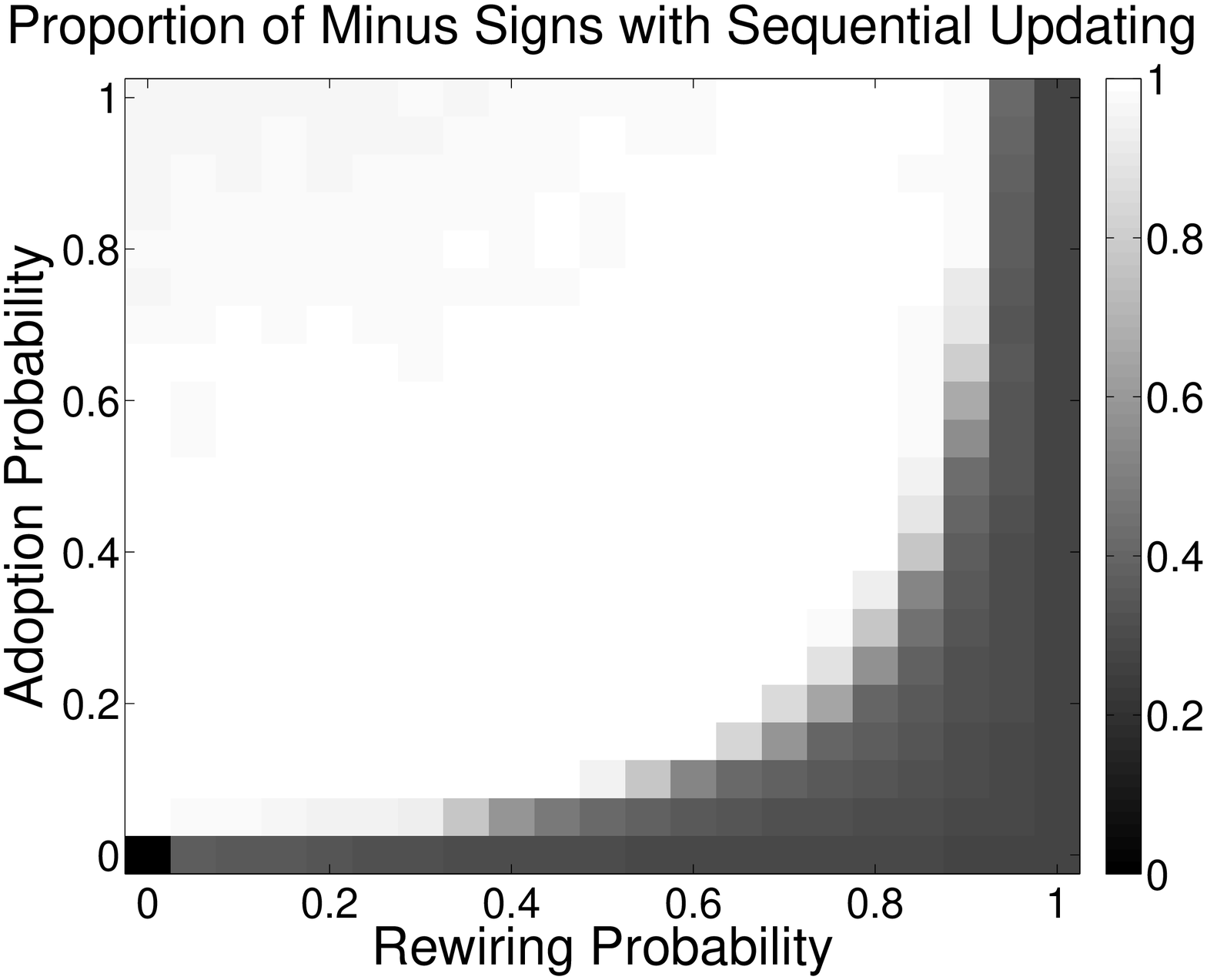}
\includegraphics[width=0.32\textwidth]{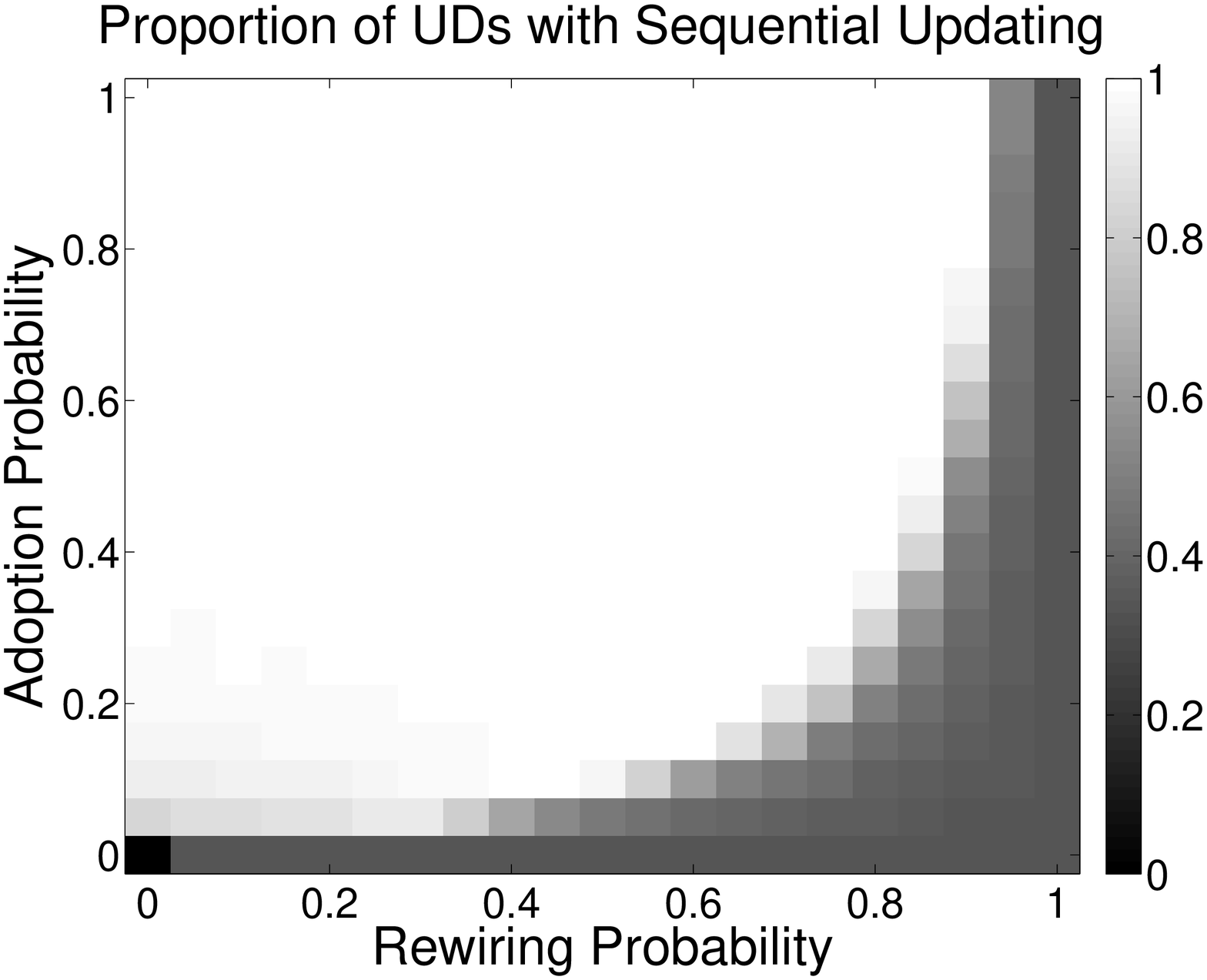}
\includegraphics[width=0.32\textwidth]{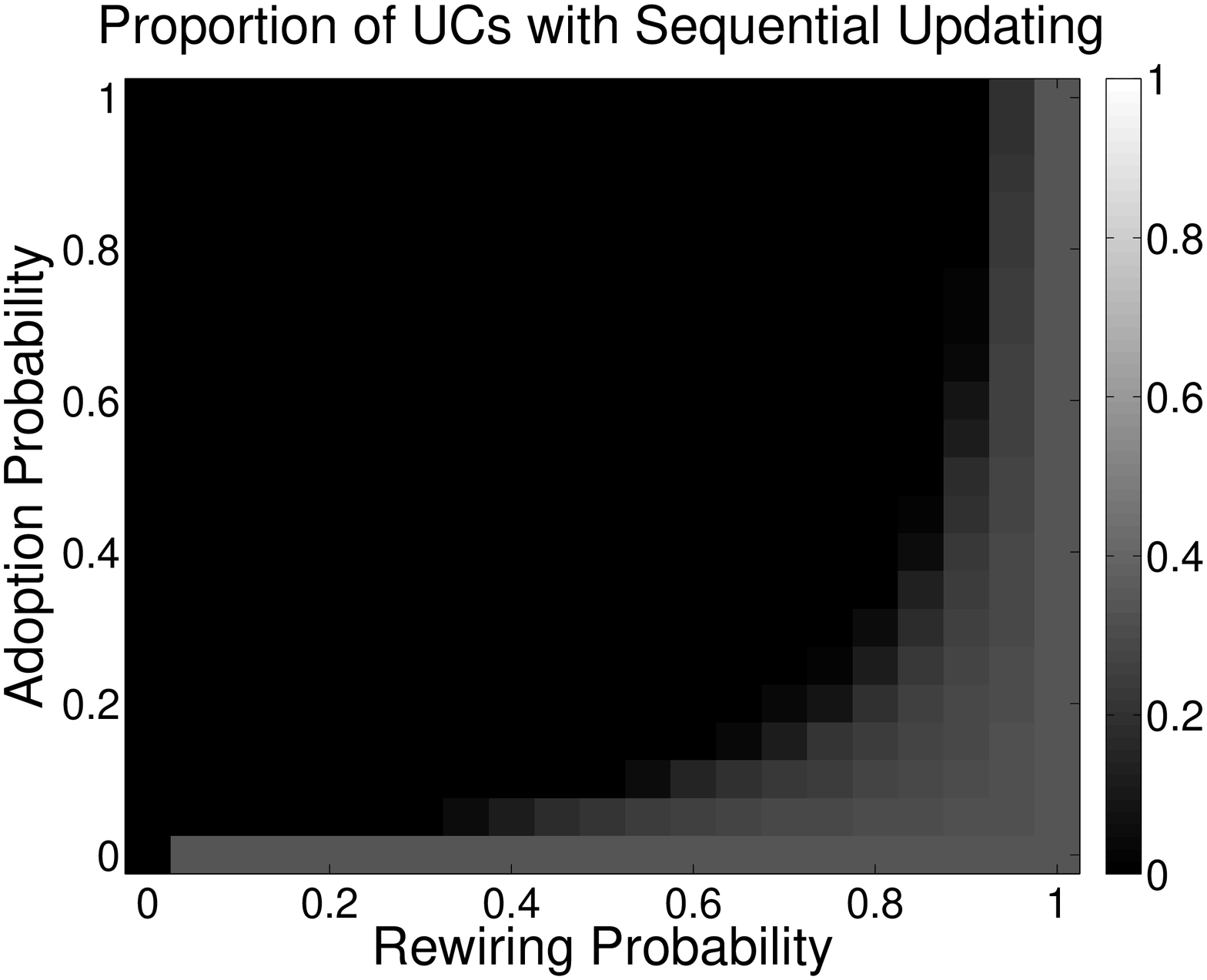}\\
\includegraphics[width=0.32\textwidth]{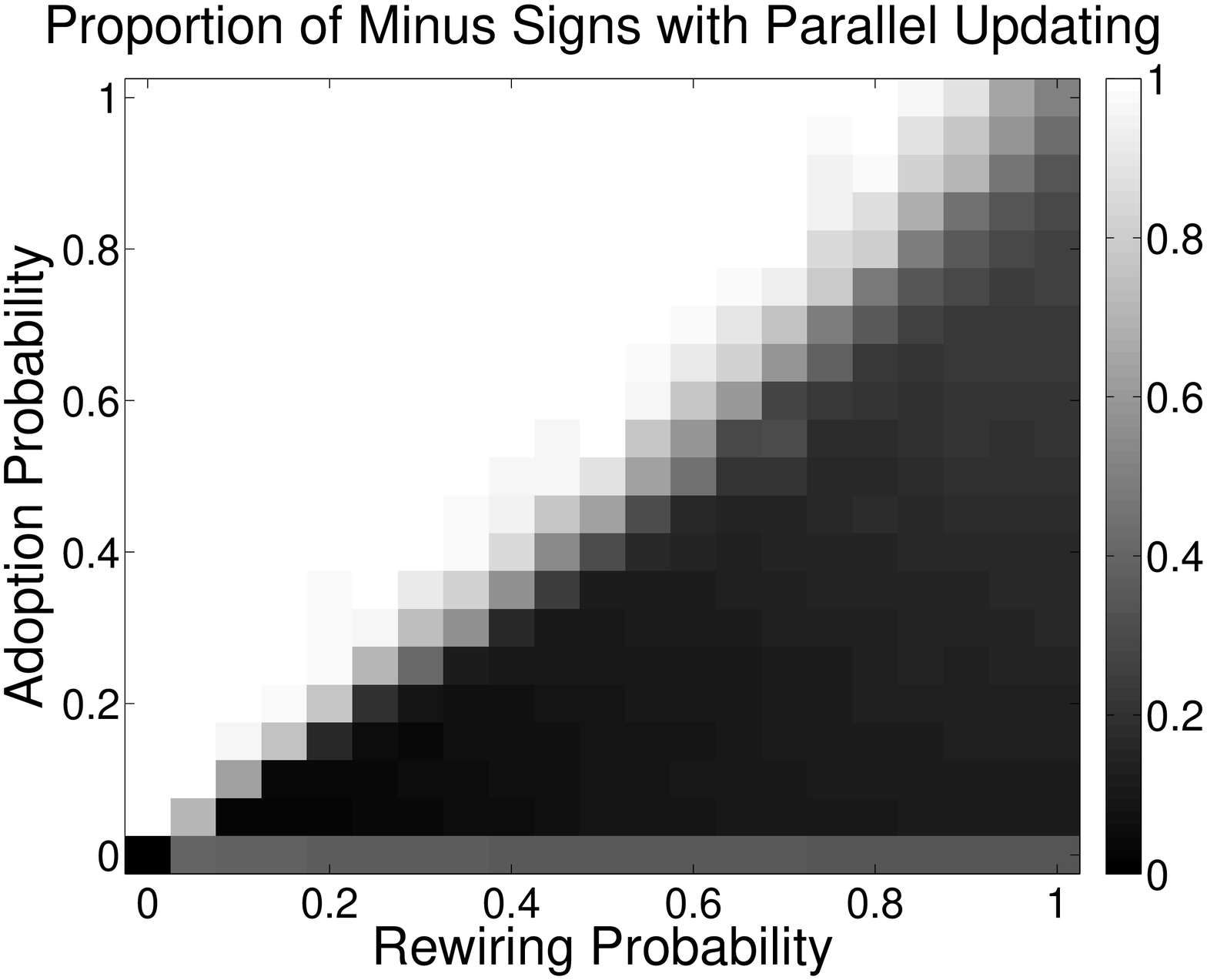}
\includegraphics[width=0.32\textwidth]{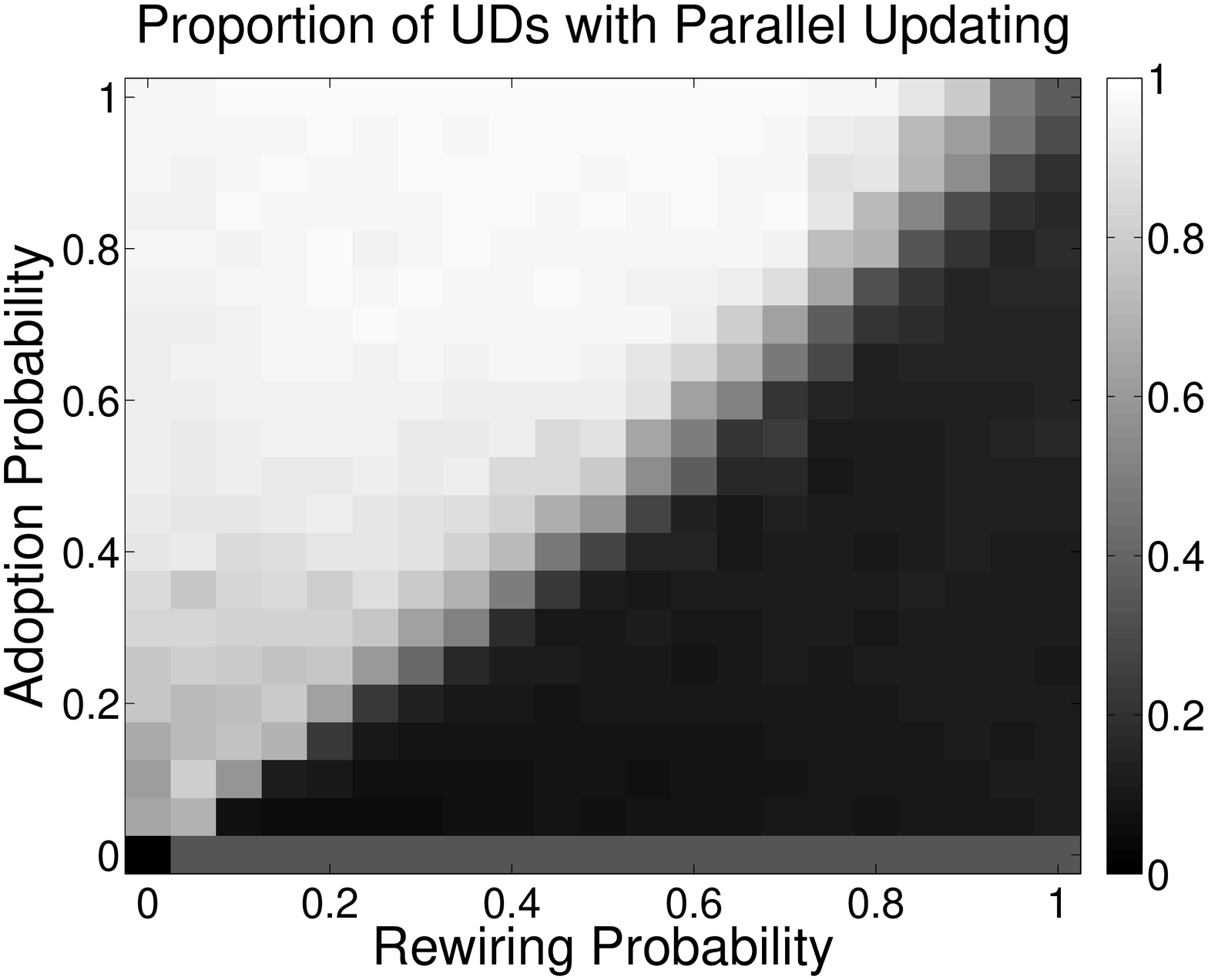}
\includegraphics[width=0.32\textwidth]{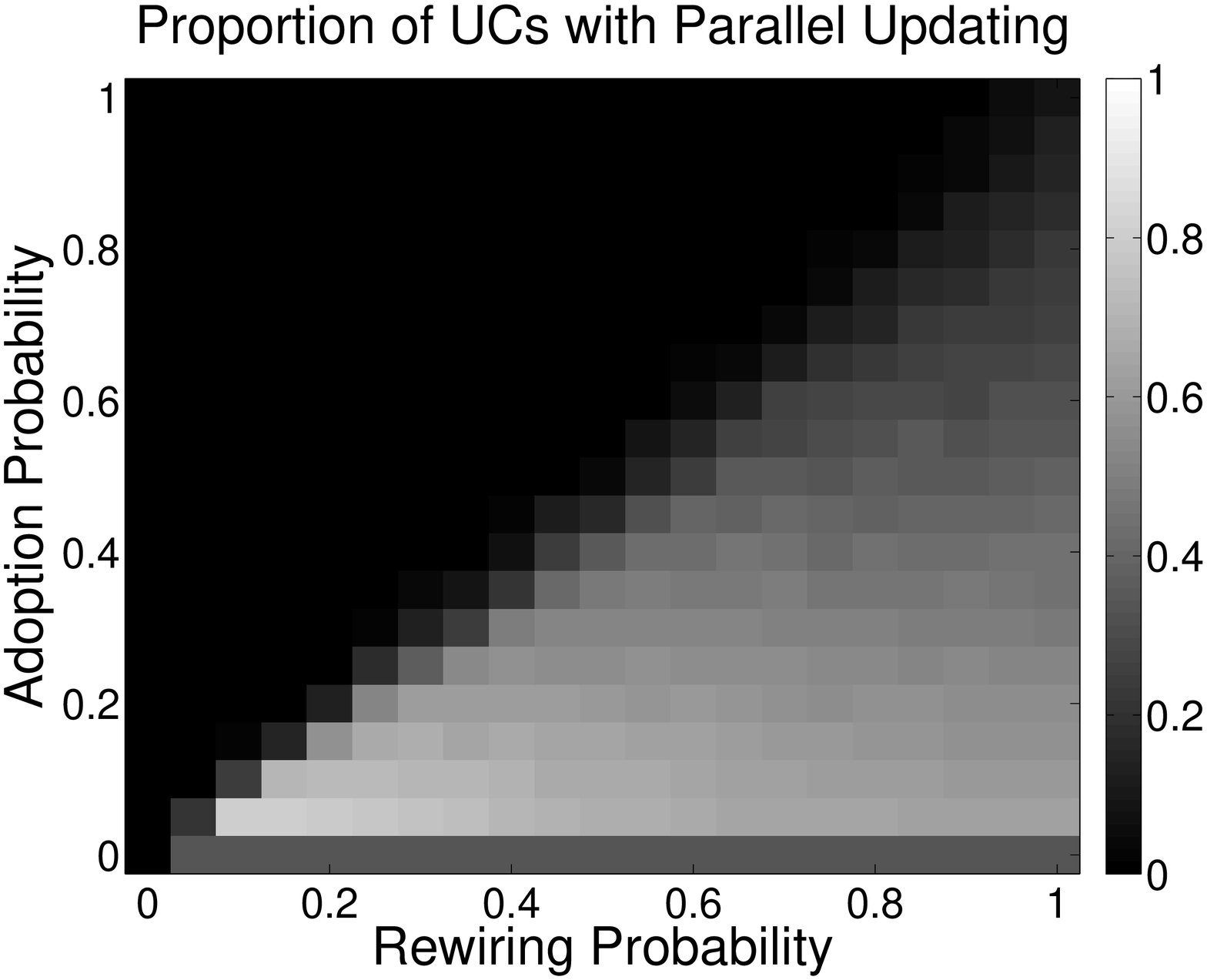}\\
\caption{Effect of the competing dynamics of adoption of strategies with higher payoffs (vertical axis) and of rewiring of stressed links (horizontal axis) on the final proportion of negative ties in the network (Left Panels), of UDs (Central Panels) and of UCs (Right Panels). Top Panels show the results for sequential updating and Lower Panels for parallel updating. In all simulations N=200. Network signs are randomly initialized with equal probability and the population is equally divided between UCs, UDs and CONDs. The probability of existence for each tie is $P_{link}=0.05$.}
\label{FixedPFlip}
\end{figure}

 Figure \ref{FixedPFlip} shows the results for the proportion of minus signs (Left Panels), unconditional defectors (Central Panels), and unconditional cooperators (Right Panels) for $P_{adopt}\in[0,1]$ and $P_{rew}\in[0,1]$ progressively changing values of both variables in steps of $0.05$. For each combination of parameters we provide the average results of 50 simulations.

Results show that, for the cooperative strategies to survive, there needs to be a relatively low invasion and a relatively high rewiring probability. This is valid for both update mechanisms and it is coherent with what observed in the literature on non-signed networks: in absence of negative ties, the rewiring mechanism limits the capacity of UDs to spread in the population.
Focusing on {\it sequential update} (Figure \ref{FixedPFlip}, Top Panels), we can observe two characteristic facts. The first is that universal cooperators may survive, but they never become dominant in the population. Indeed, their proportion never exceeds the original proportion of one third. The second observation is that cooperation survives in this setup only if $P_{rew}>>P_{adopt}$. This is in line with the positive ties literature's that shows how the scale of network update relative to the scale of strategy update is a key explanatory factor behind the chances for the evolution of cooperation (\citealt{santos2006cooperation}). Network update helps the relative clustering (it progressively eliminates negative ties with defectors) and the survival of cooperators, while frequent strategy updates provide higher importance to immediate payoffs and drive the system towards a Hobbesian destiny of no cooperation and negative links. As we have seen in the previous section, with sequential updating, the dynamics of the model tends to favor unconditional defectors that thus tend to spread in the population. This process of spread is obviously faster, the higher is $P_{adopt}$ and the rewiring mechanism is here the only mechanism that allows the survival of cooperation. 

In the case of {\it  parallel updating}, the situation is different. Cooperation can now survive also for $P_{rew} \approx P_{adopt}$. As we have seen before, the parallel update rule provides a more favorable conditions for the evolution of cooperation than sequential updating, mainly due to averaging of payoffs across multiple interactions. Averaging of payoffs and hence the increased importance of being clustered among cooperators increase the importance of the flexible character of the COND strategy. While UD still provides the best way to exploit neighbors in the short term in {\it any} environment, COND is prepared to defect and achieves at least equally good payoffs with UDs, while cooperates with cooperator neighbors and consequently earns more {\it on average} than UDs in a UD-dominated environment as well as in a COND-dominated environment. Obviously, UDs still outperform COND in a UC-dominated environment, but due to strategy updates and rewiring, such a victory is a Pyrrhic one. For these reasons, high rates of strategy update are here not that bad for conditional strategies as in the sequential update case. Emotional strategies can then act as catalysts of cooperation at least as long as $P_{rew} \geq P_{adopt}$. When $P_{adopt}$ is to high, however, strategy update favors the spread of universal defection even in the case of parallel updating.

Finally, when cooperation is supported, the proportion of UCs can extend to more than its initial value (and in some cases even above the initial sum of conditional players and unconditional cooperators) if the adoption probability is sufficiently small (with respect to rewiring). On the contrary, when both dynamic forces are strong, the proportion of UCs in the final population decreases in favor of an higher share of conditional cooperators. Indeed, a dynamic environment with high probability of rewiring associated with high probability of adoption allows the emotional strategy to extricate its power while highlighting the weakness of unconditional cooperation. On the one side, CONDs tend to diffuse since they obtain systematically higher payoffs than UDs when at the border between UCs and UDs. The intensity of the adoption rate makes it possible for CONDs to spread in the direction of universal defectors. On the other side, universal cooperators are unable to outperform UDs locally and, as a result, their survival rate decreases, being bounded below only by the fact that network rewires relatively fast. This creates clusters of both conditionals and (few) unconditional cooperators that are sustained at equilibrium.

\section{Conclusions}\label{concl}

The evolution of cooperation is one of the most puzzling problems in social sciences. In this study, we make two contributions to the resolution of the puzzle. First, building on \cite{righi2014emotional}, we make existing models more realistic by allowing negative as well as positive ties among connected agents and, in relation, we introduce and analyze the role of emotional strategies in the single-shot Prisoner's Dilemma. We show that the simple adaptive rules we defined for evolution results in the emergence of cooperation in a non-trivial way: emotional strategies act as catalysts for the success of unconditional cooperation.

Second, we analyze how this conclusion is dependent on whether we implement sequential or parallel updates in the model. With this inquiry, we follow earlier studies in complex systems that examined the importance of assuming synchronous versus real-time interactions in social dilemmas which showed that such systems might behave very differently (\citealt{huberman1993evolutionary,lumer1994synchronous}). Besides, this question is important also substantially as sequential and real-time interactions and updates are much more realistic than parallel ones. In our case, sequential updating means that single couples of individuals are selected for interactions and evolutionary and network updates immediately after. This model implementation is contrasted with parallel update, in which agents play at the same time with all their network neighbors and where average payoffs from all of their interactions determines evolutionary success. 

We provide evidence that the survival and diffusion chances of cooperation are indeed strictly limited in the sequential update rule case. Under a rather general set of parameters combinations, the parallel updating rule provides better conditions for the diffusion of cooperation. We explore the nature and range of these differences by manipulating two crucial parameters of our model: the extent of strategy updates and rewiring possibilities. While the rewiring probability should be in general higher than the strategy update one in order to find universal cooperators in the final population, the difference between the two can be small in case of parallel update, but needs to be substantial in case of sequential update.

Our results might imply that the majority of models that study the evolution of cooperation in networks or spatial settings and offer a solution for the emergence of cooperation miss an important aspect: they implicitly or explicitly assume that actions or updates are synchronous. This is in line with our parallel update rule that provides favorable conditions for the emergence of cooperation via the assistance of emotional strategies that act as catalysts in the evolutionary process. Most historical events, however, are sequential or happen in real time. As our results highlighted, the chances of cooperation are much more limited under such circumstances. This leaves the puzzle of the evolution of cooperation in case of sequential interactions still to be solved by subsequent research.

\section*{Acknowledgments}
The authors wish to thank the "Lend\"{u}let" program of the Hungarian Academy of Sciences for financial and organizational support.

\singlespacing
\bibliographystyle{apalike}
\bibliography{RighiTakacsOrleans}

\end{document}